\documentclass[apjl]{emulateapj}
\usepackage{amssymb}
\usepackage{graphicx,graphics}
\usepackage{epsfig}
\usepackage{dcolumn}
\usepackage{bm}
\usepackage{natbib}
\usepackage{times}
\usepackage{ulem}
\def\kmm#1  {{\bf [KMM:~ #1]~}}
\def\new#1 {{\bf #1 }}
\def\cut#1 {\sout{#1} }

\newcommand{\beq}{\begin{equation}}
\newcommand{\eeq}{\end{equation}}

\newcommand{\hi}{H{\sc i}}
\newcommand{\hii}{H{\sc i}~21\,cm}
\newcommand{\kms}{km~s$^{-1}$}

\newcommand{\mhi}{\ensuremath{{\rm M}_{{\rm H}{\normalfont{\tiny{\textsc{I}}}}}}}

\shorttitle{Gas mass of star-forming galaxies at $z \approx 1.3$}
\shortauthors{Kanekar et al.}
\begin{document}
\title{The gas mass of star-forming galaxies at $z \approx 1.3$}

\author{Nissim Kanekar\altaffilmark{1},
Shiv Sethi \altaffilmark{2},
K. S. Dwarakanath \altaffilmark{2}
}
\altaffiltext{1}{Swarnajayanti Fellow; National Centre for Radio Astrophysics, 
Tata Institute of Fundamental Research, Ganeshkhind, Pune - 411007, India; nkanekar@ncra.tifr.res.in}
\altaffiltext{2}{Department of Astronomy and Astrophysics, Raman Research Institute, C.V Raman Avenue, Bengaluru, India}

\begin{abstract}
We report a Giant Metrewave Radio Telescope (GMRT) search for H{\sc i} 21\,cm emission 
from a large sample of star-forming galaxies at $z \approx 1.18 - 1.34$, lying in sub-fields of 
the DEEP2 Redshift Survey. The search was carried out by co-adding (``stacking'') the 
H{\sc i} 21\,cm emission spectra of 857 galaxies, after shifting each galaxy's H{\sc i} 21\,cm 
spectrum to its rest frame. We obtain the $3\sigma$ upper limit S$_{\rm{H}\tiny{\textsc{i}}} < 2.5 
\mu$Jy on the average H{\sc i} 21\,cm flux density of the 857~galaxies, at a velocity resolution 
of $\approx 315$~km~s$^{-1}$. This yields the $3\sigma$ constraint M$_{\rm{H}\tiny{\textsc{I}}} < 2.1 
\times 10^{10} \times \left[\Delta {\rm V}/315 \rm{km/s} \right]^{1/2} \textrm{M}_\odot$ on the 
average H{\sc i} mass of the 857 stacked galaxies, the first direct constraint on the 
atomic gas mass of galaxies at $z > 1$. The implied limit on the average atomic gas mass 
fraction (relative to stars) is ${\rm M}_{\rm GAS}/{\rm M}_* < 0.5$, comparable to the 
cold molecular gas mass fraction in similar star-forming galaxies at these redshifts. 
We find that the cosmological mass density of neutral atomic gas in massive star-forming galaxies at 
$z \approx 1.3$ is $\Omega_{\rm GAS} < 3.7 \times 10^{-4}$, significantly lower than 
$\Omega_{\rm GAS}$ estimates in both galaxies in the local Universe and damped Lyman-$\alpha$ 
absorbers at $z \geq 2.2$. Massive blue star-forming galaxies thus do not appear to dominate the 
neutral atomic gas content of the Universe at $z \approx 1.3$.
\end{abstract}


\maketitle
\section{Introduction} 
\label{sec:intro}

Over the last decade, optical and infrared studies of the so-called ``deep fields'' 
\citep[e.g.][]{giavalisco04,scoville07,newman13} have revolutionized our 
understanding of galaxy formation and evolution. Detailed studies of the cosmic star formation 
history have shown that the comoving star formation rate (SFR) density rises towards the end of 
epoch of reionization and peaks in the redshift range $z \approx 1-3$, before declining by 
an order of magnitude from $z\approx 1$ to today \citep[e.g.][]{lefloch05,hopkins06,bouwens09}. 
Indeed, the range $z \approx 1-3$ is often referred to as the ``epoch of galaxy assembly'', as 
roughly half of today's stellar mass density was formed during this period \citep[e.g.][]{reddy08,marchesini09}. 
It is also known that the nature of the galaxies undergoing star formation evolves significantly from 
$z\approx 2$ to today, via cosmic downsizing \citep{cowie96}: the SFR density at $z \sim 2$ 
is dominated by massive galaxies with high SFRs, while that at $z \approx 0$ mostly arises in 
lower-mass systems with lower SFRs \citep[e.g.][]{lefloch05,seymour08}. A tight relation 
(the ``main sequence'') has been found between the specific SFR (sSFR) and the stellar mass of 
most high-$z$ star-forming galaxies \citep[][]{noeske07}, while a small fraction, the starbursts, 
undergo far more efficient star formation than main-sequence galaxies of the same stellar mass, 
with sSFRs larger by an order of magnitude \citep[e.g.][]{rodighiero11}.

The above remarkable recent progress in studies of stars in high-$z$ galaxies has not 
been mirrored in studies of the other major constituent of galaxies, the neutral gas, 
the fuel for the star formation. In the local Universe, this gas is best probed with emission 
studies in the \hii\ line. Unfortunately, the weakness of the \hii\ transition has meant that 
the highest redshift at which it has so far been detected in emission is $z \approx 0.25$ 
\citep[e.g.][]{catinella15}. 
In the case of the molecular component,
 studies of CO in emission at $z \approx 1.5 - 3$ have indeed found evidence for massive 
reservoirs of molecular gas (${\rm M}_{\rm H_2} \gtrsim 10^{10} {\rm M}_\odot$, comparable to the stellar 
mass) in star-forming galaxies \citep[e.g.][]{daddi10b,tacconi10,tacconi13}. Unfortunately, CO is only 
a tracer of the bulk of the molecular gas, and the conversion factor from the CO line luminosity 
to molecular gas mass is known to depend on galaxy type, varying by a factor of 
$\approx 5$ between starburst and spiral galaxies, and by larger factors in 
low-metallicity galaxies \citep[e.g.][]{genzel12,carilli13}.


Overall, the situation today is that little is known about the neutral gas, especially the 
atomic component, in high-$z$ star-forming galaxies. Unfortunately, even very deep integrations with 
today's best radio telescopes will only be able to detect \hii\ emission from even the most massive galaxies 
($\mhi \gtrsim few \times 10^{10} {\rm M}_\odot$) out to relatively low 
redshifts, $z \lesssim 0.5$ \citep[e.g.][]{verheijen07,fernandez13}; detecting \hii\ emission from 
individual galaxies at $z \gtrsim 1$ will need the large collecting area of the Square Kilometre Array. 
However, information can be obtained about the {\it average} gas properties of high-$z$ galaxies by 
co-adding (``stacking'') the \hii\ emission signals of a large number of galaxies with known redshifts 
that lie within the primary beam of the radio telescope but that are too faint to be detected individually 
\citep{zwaan00,chengalur01}. Such \hii\ stacking can be used to infer the average \hi\ mass of the 
galaxies and the cosmological mass density in neutral gas in these systems \citep[e.g.][]{lah07,chang10,delhaize13,rhee13}.
One can also determine average properties of the galaxy sample, including the dependence 
of the gas mass on various attributes (e.g. SFR, stellar mass, environment, and redshift),
the average Tully-Fisher relation, etc \citep[e.g.][]{fabello11,fabello12,meyer16}. 


Stacking of the \hii\ spectra of individual galaxies with known redshifts, obtained using radio 
interferometry, has so far only been carried out at low redshifts, $z \lesssim 0.4$ \citep[e.g.][]{lah07,lah09,rhee13}.
This has been extended to slightly higher redshifts, $z \approx 0.8$, via cross-correlation of 
single-dish \hii\ intensity maps with deep optical images \citep{chang10,masui13}. In this {\it Letter}, 
we report Giant Metrewave Radio Telescope (GMRT) results from the first \hii\ stacking experiment at 
$z > 1$, obtaining tight constraints on the average gas mass of star-forming galaxies at $z \approx 1.3$.

\section{The DEEP2 Survey fields}
\label{sec:sample}

The stacking approach critically requires a large sample of galaxies with accurately 
known redshifts in the target field. This is a significant restriction when 
selecting fields for searches for \hii\ emission at $z \approx 1-1.5$, as this is the 
``redshift desert'' where the bright optical lines (e.g. H$\alpha$, O{\sc ii}, H$\beta$) 
that are typically used to identify galaxies are redshifted to wavelengths
$\gtrsim 7500$~\AA, into regions full of night-sky lines. Most spectroscopic galaxy 
samples in deep fields are hence dominated by galaxies at $z \lesssim 0.7$, or at 
$z \gtrsim 2$. The DEEP2 Galaxy Redshift Survey is unique in this regard as its 
redshift coverage extends out to $z \approx 1.4$ \citep{davis03,newman13}, due to the use 
of the high spectral resolution ($R = 6000$) mode of the DEIMOS spectrograph on 
the Keck-II Telescope, allowing the clear identification of the O{\sc ii}$\lambda 3727$ 
doublet out to $z \approx 1.4$. The DEEP2 Survey covers four fields with a total areal 
coverage of 2.8~deg.$^2$, and has yielded accurate redshifts for 38,000 massive galaxies 
in the redshift range $0.75 \lesssim z \lesssim 1.4$ \citep{newman13}. A few thousand 
galaxies lie at $z \approx 1.2-1.4$, for which the \hii\ line is redshifted to 
$\approx 590-650$~MHz, i.e. into the GMRT 610~MHz band. This picks out the DEEP2 
Survey fields as the ideal targets for a GMRT search for \hii\ emission at $z \approx 1.3$.
Three of the DEEP2 fields are of size $120' \times 30'$, each consisting of three 
sub-fields of size $36' \times 30'$, while the fourth, the Extended Groth Strip (EGS), is of 
size $120' \times 15'$. The size of the sub-fields of the first three DEEP2 fields are 
well-matched to the size of the GMRT primary beam at 610~MHz, which has a full width at half
maximum (FWHM) of 
$\approx 43'$. Hence, although the EGS has excellent ancillary multi-wavelength 
information, we chose to focus on the DEEP2 sub-fields 21, 22, 31/32 and 32/33 
\citep[see][]{newman13}, all with full spectroscopic coverage, for the present 
GMRT pilot survey.

\section{Observations and Data Analysis}
\label{sec:data}

\begin{figure}[t!]
\centering
\includegraphics[scale=0.4]{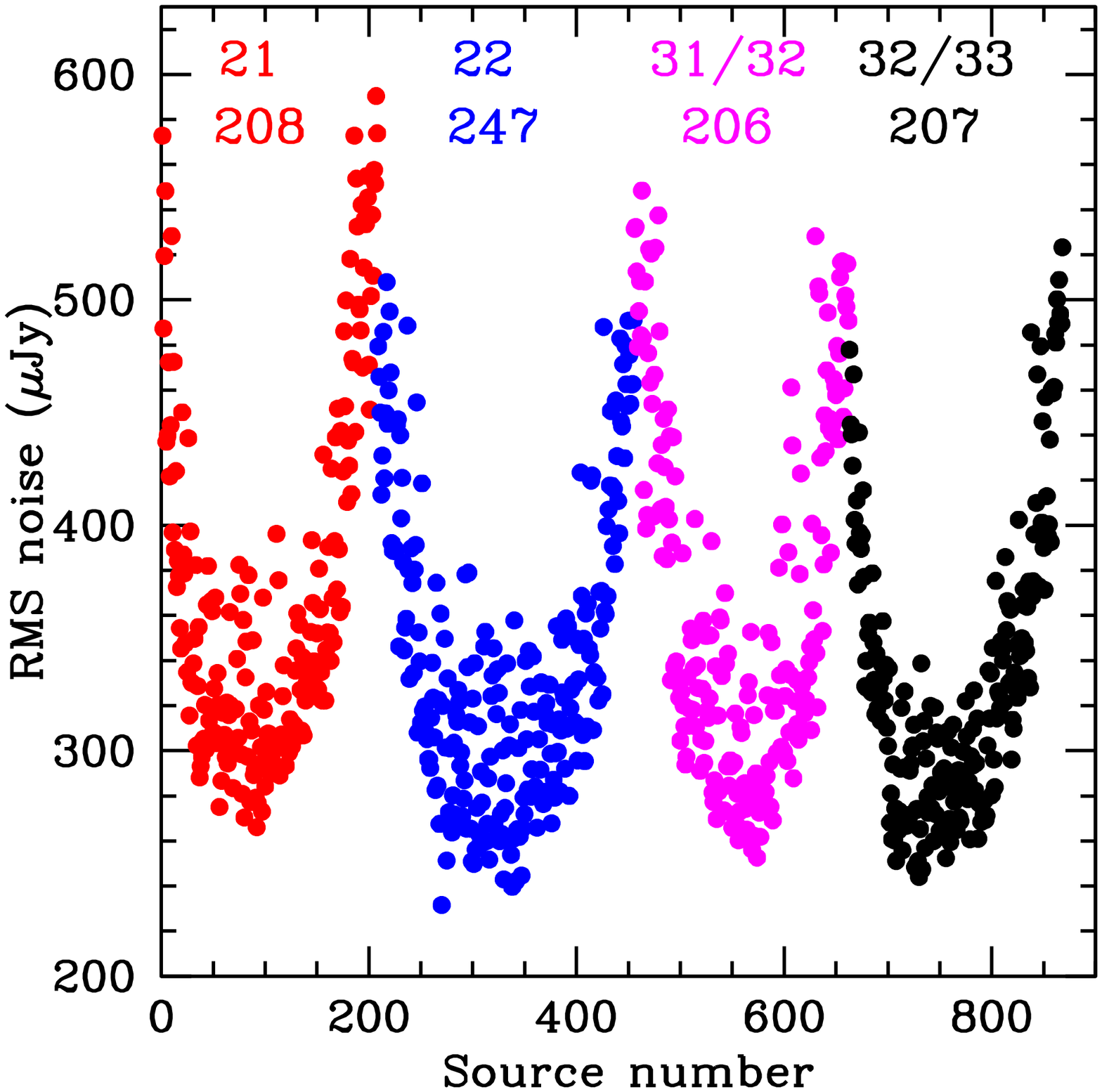}
\caption{The spectral RMS noise values for the 868 target galaxies, after correcting for
their location within the GMRT primary beam. The abscissa is an arbitrary source number;
each field is labeled by the GMRT pointing center (DEEP2 sub-fields 21, 22, 31/32, 
32/33) and the number of galaxies included in the GMRT spectral coverage.
\label{fig:rms}}
\end{figure}

We used the GMRT 610~MHz receivers to observe the four DEEP2 sub-fields in 2011 November
(sub-fields 21 and 22, with pointing centers of RA=16h48m00.0s, Dec=+34d56'00.0" and 
RA=16h51m00.0s, Dec=+34d56'00.0", respectively) and 2012 October and November (pointing 
midway between sub-fields 31 and 32, and 32 and 33, with pointing centers of 
RA=23h28m00.0s, Dec=+00d09'00.0" and RA=23h32m00.0s, Dec=+00d09'00.0", respectively).
A bandwidth of 33.33 MHz, sub-divided into 512 channels, was used for all observations, 
with 30 antennas, and the GMRT Software Backend as the correlator. The frequency coverage
was $601-634.33$~MHz in 2011 and $621-654.33$~MHz in 2012, implying a redshift coverage 
of $\approx 1.24 - 1.36$ and $ 1.18-1.28$, respectively, with a velocity resolution 
of $\approx 32$~km~s$^{-1}$. The different frequency settings were used to test the 
possibility that radio frequency interference (RFI) might restrict the sensitivity in 
different parts of the GMRT 610~MHz band. Observations of 3C286 and 3C48 were used to 
calibrate the flux density scale, and of 0022+002 and 1609+266 to calibrate the 
antenna gains and bandpasses. The total on-source times on each field ranged from 
500\,minutes (field 32) to 804\,minutes (field 22).

The data were analysed in ``classic'' {\sc aips}, using standard procedures. For each field, 
after initial data editing and calibration of the antenna bandpasses and gains, the visibility 
data were averaged to produce a dataset of spectral resolution $\approx 1$~MHz. Accurate
antenna-based gains were then derived via a self-calibration procedure, first iterating between
3-D imaging and phase-only self-calibration for a few rounds, followed by amplitude-and-phase
self-calibration, 3-D imaging, and data editing. The procedure was repeated until the 
continuum image did not improve on additional self-calibration; the continuum root-mean-square (RMS) 
noise near the center of the primary beam is $\approx 23 - 35 \mu$Jy in three of the GMRT 
pointings, but $\approx 100 \mu$Jy for sub-field 22, probably due to dynamic range issues due 
to bright sources at or below the half-power point of the primary beam.

For each field, the task {\sc uvsub} was then used to subtract the final image from the calibrated 
spectral-line visibilities, after which any residual continuum was subtracted via a linear fit 
to each visibility spectrum using the task {\sc uvlin}. The residual visibilities were then 
shifted to the heliocentric frame, using the task {\sc cvel}. Polyhedral imaging was then used, 
with natural weighting, to produce wide-field spectral cubes, extending beyond the 
FWHM of the primary beam. The angular resolution of the final 
spectral cubes ranges from $\approx 9.7'' \times 7.8''$ to $\approx 7.2'' \times 5.5''$, 
corresponding to a spatial size of $\gtrsim 60 \, {\rm kpc} \times 50$~kpc at 
$z = 1.3$\footnote{We assume a standard $\Lambda$-cold-dark-matter ($\Lambda$CDM) cosmology, 
with $H_0 = 67.8$~km~s$^{-1}$~Mpc$^{-1}$, $\Omega_m = 0.304$ and $\Omega_\Lambda = 0.696$ 
\citep{planck15}.}. 

\section{Stacking the H{\sc i} 21\,cm spectra}
\label{sec:stacking}

\begin{figure*}[t!]
\centering
\includegraphics[scale=0.4]{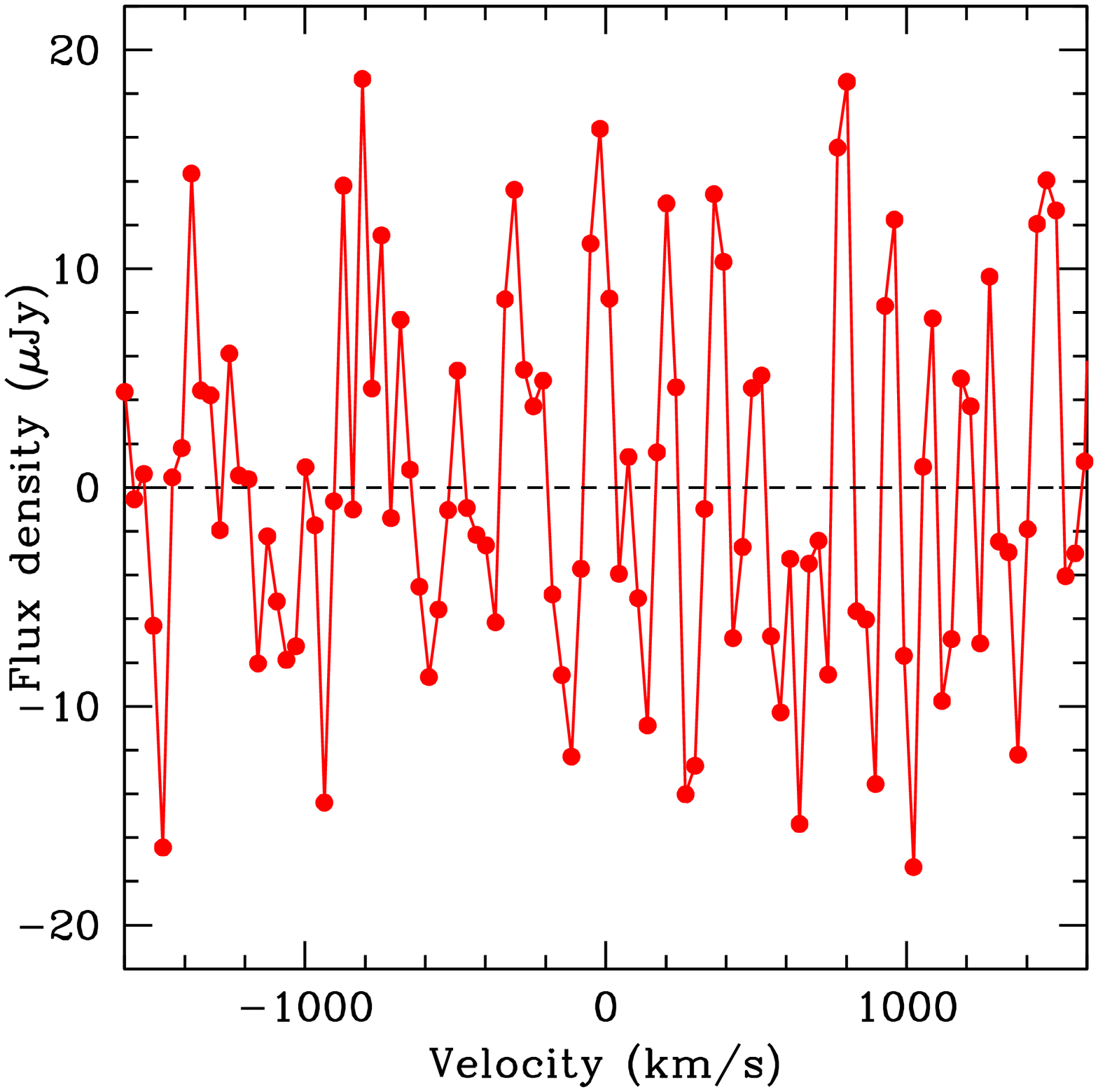}
\includegraphics[scale=0.4]{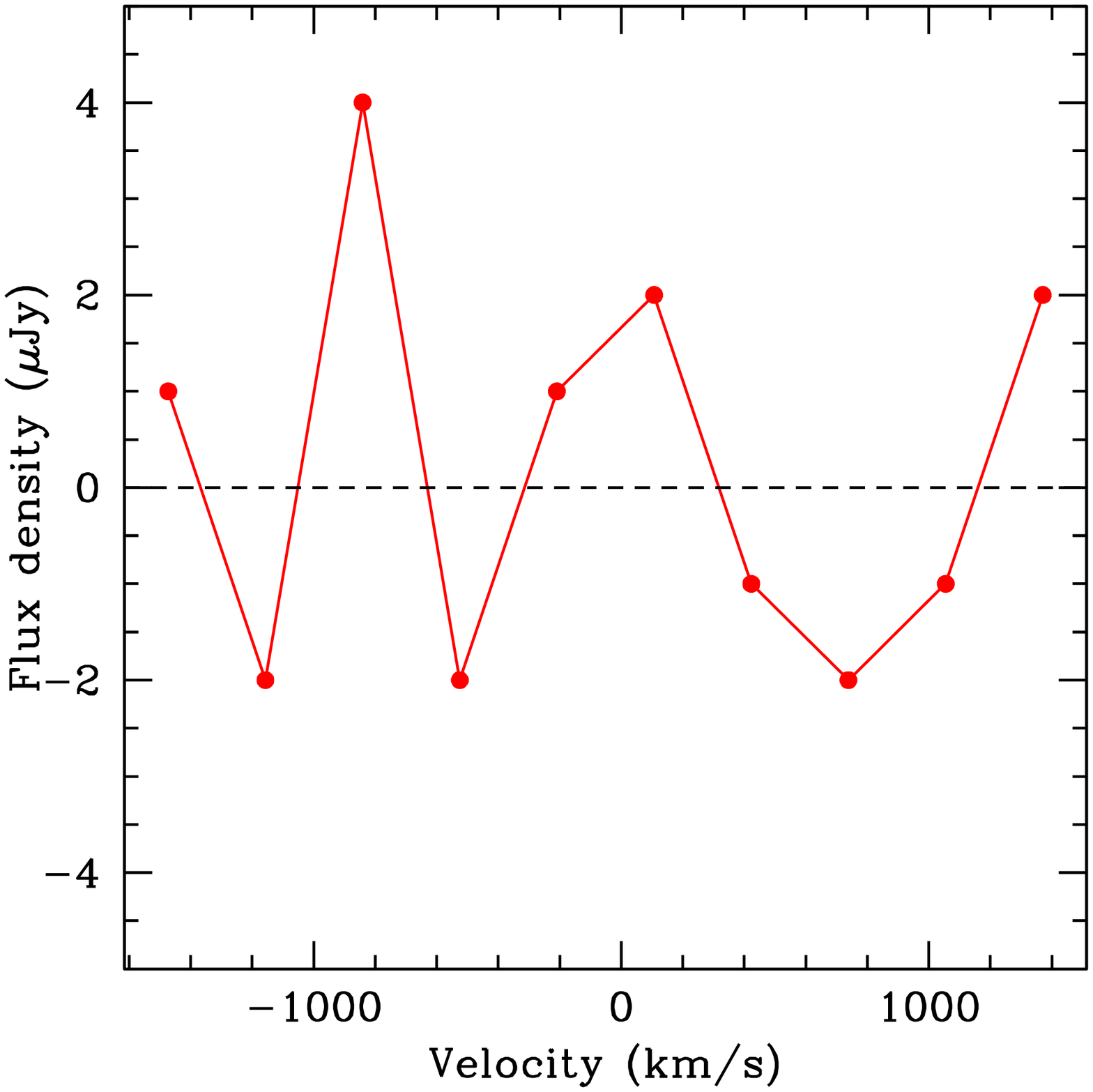}
\caption{The final GMRT \hii\ spectra, after stacking the \hii\ spectra of 857 DEEP2 galaxies,
at [A]~the original velocity resolution of $\approx 32$~km~s$^{-1}$ and [B]~a velocity 
resolution of $\approx 315$~km~s$^{-1}$, after smoothing and resampling. No evidence for 
a statistically significant emission feature is apparent in either spectrum.
\label{fig:spc}}
\end{figure*}

We stacked the \hii\ spectra of all DEEP2 galaxies lying within the FWHM of the GMRT primary beam 
and with accurately known redshifts \citep[redshift quality $\geq 3$ in the DEEP2 catalog;][]{newman13} that shift
the \hii\ line into the observed band for each target field. We initially excluded 20 channels 
from each edge of the observing band, where the sensitivity is significantly lower than in the 
band center. Next, in order to have sufficient spectral baseline around any putative 
stacked spectral feature, we only considered galaxies with redshifts $\geq 1500$~km~s$^{-1}$ 
from the edges of the retained frequency range. This yielded $206-247$ galaxies in the four 
GMRT pointings, with a total of 868~galaxies. \hii\ spectra were obtained by taking a cut through 
the different spectral cubes at the location of each galaxy. A second-order baseline was fit to 
each spectrum and subtracted out to remove any residual curvature, and each spectrum was then 
scaled to account for the location of the galaxy within the GMRT primary beam. The distribution 
of spectral RMS noise values per $\approx 32$~km~s$^{-1}$ channel for the different spectra, 
after the above primary beam correction, is shown in Fig.~\ref{fig:rms}; the RMS noise at the 
center of each field is $\approx 240-260 \mu$Jy per $\approx 32$~km~s$^{-1}$ channel.
  
Each spectrum was tested for gaussianity, via both the Kolmogorov-Smirnov rank-1 (K-S) and 
Anderson-Darling (A-D) tests; 11 spectra found to have p-values $< 0.003$ in the A-D test 
were excluded from the stacking (note that retaining these spectra does not significantly affect
our results). The \hii\ spectrum for each retained galaxy was next shifted to its rest 
frame and then interpolated onto a single velocity axis, before the spectra of each field 
were optimally co-added with weights determined from their RMS noise values (after the primary 
beam correction; see Fig.~\ref{fig:rms}) to produce the final stacked spectrum for each field. 
These four spectra were then co-added, again with appropriate weights based on their RMS noise 
values, and a second-order baseline subtracted out from the result, to produce the final stacked 
\hii\ spectrum. This is shown in Fig.~\ref{fig:spc}[A] and has an RMS noise of $\approx 8.7 \mu$Jy 
per $\approx 32$~km~s$^{-1}$. 

The signal-to-noise ratio in a search for line emission is maximized on smoothing the 
spectrum to a resolution equal to the line FWHM. We hence searched for \hii\ emission in 
the stacked spectrum after smoothing to, and resampling at, a range of velocity resolutions 
from $\approx 100 - 500$~km~s$^{-1}$; no statistically significant feature was detected at 
any resolution. Fig.~\ref{fig:spc}[B] shows the spectrum smoothed to, and resampled at, a 
resolution of $\approx 315$~\kms; this has an RMS noise of $\approx 2.1 \mu$Jy and shows 
no evidence for \hii\ emission. 

The $3\sigma$ upper limit on the strength of \hii\ emission in the final spectrum at a resolution 
of $\approx 315$~\kms\ can be used to derive a limit on the average \hi\ mass of the 857 stacked 
galaxies. However, our estimate of the RMS noise on this spectrum is based on only 10 points.
We hence used a bootstrap approach to obtain a more reliable estimate of the RMS noise on the 
stacked spectrum, by stacking the 857 spectra after randomizing the galaxy redshifts; this 
was done for 100 independent realizations and yielded an average RMS noise of $\approx 2.5 \mu$Jy
per $\approx 315$~\kms\ channel. This estimate of the RMS noise will be used in the later analysis.

\section{Discussion and Summary}
\label{sec:discuss}

\begin{figure}[t!]
\centering
\includegraphics[scale=0.4]{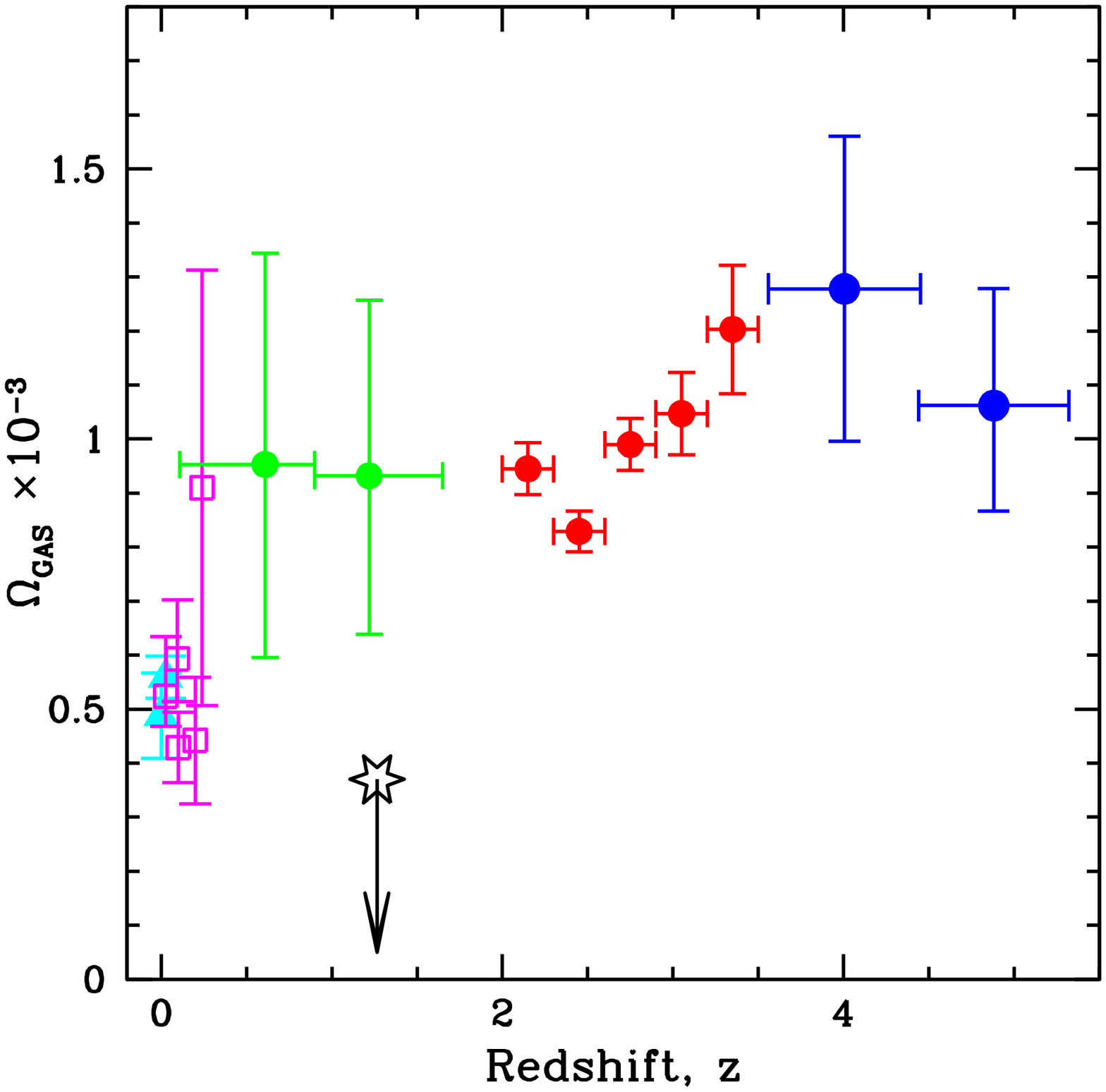}
\caption{The cosmological mass density in neutral gas $\Omega_{\rm GAS}$ plotted as a function 
of redshift. The filled triangles are from \hii\ emission surveys \citep{zwaan05b,martin10},
the open squares from \hii\ emission stacking at low redshifts \citep{lah07,delhaize13,rhee13},
and the filled circles from absorption-selected galaxies, Mg{\sc ii} absorbers at $z \approx 0.7-1.2$ 
\citep{rao06} and damped Lyman-$\alpha$ absorbers at $z \geq 2.2$  \citep{noterdaeme12,crighton15}.
The open star shows our new GMRT result, for blue star-forming galaxies at $z \approx 1.265$;
it is clear that this is significantly lower than the $\Omega_{\rm GAS}$ estimates at both $z = 0$ 
and $z \geq 2.2$ [see also \citet{neeleman16}].
\label{fig:omega}}
\end{figure}

The DEEP2 galaxy sample is selected at R-band, with a limiting R-magnitude of 24.1 
\citep{newman13}. For the median redshift of $z_{\rm med} = 1.265$ of our 857 galaxies,
this corresponds to selection at a rest-frame wavelength of $\approx 2950$~\AA. As pointed out by 
\citet{weiner09}, the high-$z$ DEEP2 sample is hence highly biased towards star-forming 
galaxies in the blue cloud of the galaxy color-magnitude diagram \citep[e.g.][]{willmer06}. 
Our target galaxies are thus expected to predominantly be blue star-forming galaxies.
\citet{cooper06} find that $< 2$\% of DEEP2 galaxies at all redshifts lie in clusters, 
with $\approx 67$\% in the field and $\approx 30$\% in small groups; the fraction of 
cluster galaxies is even smaller for blue, star-forming galaxies. Ram-pressure stripping of the 
\hi\ in our target galaxies \citep[as observed in nearby clusters, e.g.][]{giovanelli83,verheijen07} 
is thus unlikely to be a serious issue.

For the assumed $\Lambda$CDM cosmology and the median redshift $z_{\rm med} = 1.265$ 
of our stacked galaxies, our non-detection of \hii\ emission in the stacked spectrum of 
Fig.~\ref{fig:spc}[B] implies the $3\sigma$ upper limit $\mhi < 2.1 
\times 10^{10} {\rm M}_\odot \times \left[\Delta {\rm V}/315 \right]^{1/2}$ on the 
average \hi\ mass of the 857 stacked galaxies, where we have assumed a Gaussian line shape 
with FWHM~$=315$~km~s$^{-1}$. Note that the large spatial extent 
of the GMRT synthesized beam ($> 60$~kpc~$\times 50$~kpc at $z \approx 1.3$) implies that 
it is very unlikely that we are resolving out the \hii\ emission. This is the first direct 
constraint on the atomic gas mass of star-forming galaxies at $z > 1$.

Stellar mass estimates are available for galaxies in two of our target fields (22 
and 32) from K-band imaging \citep{bundy06}, with an average stellar mass of ${\rm M}_* 
\approx 5.1 \times 10^{10} {\rm M}_\odot$. Assuming that our other target galaxies have 
the same average stellar mass then gives an atomic gas mass fraction (relative to stars, 
and after including a factor of 1.32 to account for the contribution of helium) of 
${\rm M}_{\rm GAS}/{\rm M}_* < 0.5$, at $3\sigma$ significance. Note that \citet{tacconi13} 
used CO studies of star-forming galaxies at $z \approx 1.2$ in the EGS to infer an average cold 
molecular gas fraction relative to stars of ${\rm M}_{\rm Mol}/{\rm M}_* \approx 0.33$, 
comparable to our upper limit on the atomic gas fraction. 
Our results thus indicate that neutral gas in massive star-forming galaxies at $z \approx 1.3$ 
is likely to be mostly in the molecular phase \citep[e.g.][]{tacconi13,lagos14}.

Since the DEEP2 galaxies were selected uniformly in R-band as a magnitude-limited sample 
\citep{newman13}, our 
non-detection of \hii\ emission can be used to constrain the cosmological density of 
neutral gas in star-forming galaxies at $z \approx 1.265$. The FWHM of the GMRT primary beam 
at 627.1~MHz (corresponding to the redshifted \hii\ line frequency for $z = 1.265$) is $43.2'$,
implying a comoving area of $50.3$~Mpc~$\times 50.3$~Mpc. Our stacking analysis 
first excluded 20 edge channels, and then galaxies with redshifts $\leq 1500$~km~s$^{-1}$ 
from the retained channels; this implies an effective bandwidth of $\approx 23$~MHz 
centered at 627.1~MHz, i.e. a comoving line-of-sight distance of 177.8~Mpc. The effective 
comoving volume of a single GMRT pointing is hence $V \approx 3.54 \times 10^5$~Mpc$^3$.
The upper limit to the cosmological mass density of \hi\ in star-forming galaxies 
is then $\Omega_{{\rm H}\tiny{\textsc{I}} {\rm , SF}} = {\rm M}_{{\rm H}{\tiny{\textsc{I}}}} \times N/V$, where 
$N = 851/4$ is the average number of DEEP2 galaxies in a single GMRT pointing. This yields
$\Omega_{{\rm H}{\tiny{\textsc{I}}} {\rm , SF}} < 2.8 \times 10^{-4}$. To obtain the cosmological 
mass density of neutral gas in star-forming galaxies, $\Omega_{\rm GAS, SF}$, we include a 
factor of 1.32 for helium to obtain $\Omega_{\rm GAS, SF} < 3.7 \times 10^{-4}$. 

Fig.~\ref{fig:omega} plots $\Omega_{\rm GAS}$ versus redshift from a variety of measurements 
at different redshifts. 
The figure shows that our $3\sigma$ upper limit to $\Omega_{\rm GAS, SF}$ is significantly lower 
(by a factor $\gtrsim 1.5$) than the measured $\Omega_{\rm GAS}$ at both $z \approx 0$ and 
$z \geq 2.2$. It thus appears that blue star-forming galaxies do not dominate the 
cosmological gas mass density in neutral atomic gas at $z \approx 1.3$.

We note that it is possible that the most \hi-rich galaxies at a given redshift do not dominate 
the star formation activity. For example, the ``\hi\ monsters'' of \citet{lee14}, with 
\hi\ masses $> 3 \times 10^{10}\: {\rm M}_\odot$ at $z \approx 0.04-0.08$, have SFRs 
$\approx 0.5 - 5 \: {\rm M}_\odot$/yr. Our results thus do not rule out the possibility that 
$\Omega_{\rm GAS}$ at $z \approx 1.3$ is dominated by massive, low-surface brightness 
galaxies with low SFRs. 

In summary, we have used the GMRT to carry out a search for \hii\ emission at $z \approx 1.265$
by stacking the \hii\ emission signals from 857 star-forming galaxies in four DEEP2 survey 
fields. Our non-detection of \hii\ emission in the stacked \hii\ spectrum yields the $3\sigma$ 
upper limit $\mhi < 2.1 \times 10^{10} \times \left[\Delta {\rm V}/315 {\rm km/s} \right] M_\odot$ 
on the average \hi\ mass of galaxies in our sample. Comparing this to the average stellar 
mass of DEEP2 galaxies yields the $3\sigma$ upper limit ${\rm M}_{\rm GAS}/{\rm M}_* < 0.5$ 
on the atomic gas mass fraction relative to stars. This is similar to the measured average cold 
molecular gas mass fraction in star-forming galaxies at $z \approx 1.2$, suggesting that 
the neutral gas in these galaxies may be mostly in the molecular phase. Finally, we obtain 
the limit $\Omega_{\rm GAS, SF} < 3.7 \times 10^{-4}$ on the cosmological mass density of 
neutral gas in star-forming galaxies at $z \approx 1.3$. This is significantly lower 
than estimates of $\Omega_{\rm GAS}$ at both $z = 0$ and $z \geq 2.2$, indicating that 
massive blue star-forming galaxies do not dominate the gas content of the Universe during the 
epoch of galaxy assembly.

\acknowledgments
We thank Kate Rubin, Ben Weiner, Xavier Prochaska and Claudia Lagos for discussions,
and Kevin Bundy for providing the stellar mass estimates. We also thank the 
GMRT staff who have made these observations possible. The GMRT is run by the National Centre for 
Radio Astrophysics of the Tata Institute of Fundamental Research. NK acknowledges support from the 
Department of Science and Technology via a Swarnajayanti Fellowship (DST/SJF/PSA-01/2012-13).

\bibliographystyle{apj}

\end{document}